

\font\titlefont = cmr10 scaled\magstep 4
\font\sectionfont = cmr10
\font\littlefont = cmr5
\font\eightrm = cmr8

\magnification = 1200

\global\baselineskip = 1.2\baselineskip
\global\parskip = 4pt plus 0.3pt
\global\abovedisplayskip = 18pt plus3pt minus9pt
\global\belowdisplayskip = 18pt plus3pt minus9pt
\global\abovedisplayshortskip = 6pt plus3pt
\global\belowdisplayshortskip = 6pt plus3pt


\def\endignore{}
\def\ignore #1\endignore{}

\newcount\dflag
\dflag = 0


\def\monthname{\ifcase\month
\or Jan \or Feb \or Mar \or Apr \or May \or June%
\or July \or Aug \or Sept \or Oct \or Nov \or Dec
\fi}

\def\timestring{{\count0 = \time%
\divide\count0 by 60%
\count2 = \count0
\count4 = \time%
\multiply\count0 by 60%
\advance\count4 by -\count0
\ifnum\count4 < 10 \toks1 = {0}
\else \toks1 = {} \fi%
\ifnum\count2 < 12 \toks0 = {a.m.}
\else \toks0 = {p.m.}
\advance\count2 by -12%
\fi%
\ifnum\count2 = 0 \count2 = 12 \fi
\number\count2 : \the\toks1 \number\count4%
\thinspace \the\toks0}}



\def\endtitle{}
\def\title#1\endtitle{\vskip.5in\titlefont
\global\baselineskip = 2\baselineskip
#1\vskip.4in
\baselineskip = 0.5\baselineskip\rm}

\def\endauthors{}
\def\authors#1\endauthors{#1}

\def\endabstract{}
\def\abstract#1\endabstract{\vskip .3in%
\centerline{\sectionfont\bf Abstract}%
\vskip .1in
\noindent#1}

\newcount\nsection
\newcount\nsubsection

\def\section#1{\global\advance\nsection by 1
\nsubsection=0
\bigskip\noindent\centerline{\sectionfont \bf \number\nsection.\ #1}
\bigskip\rm\nobreak}

\def\subsection#1{\global\advance\nsubsection by 1
\bigskip\noindent\sectionfont \sl \number\nsection.\number\nsubsection)\
#1\bigskip\rm\nobreak}


\def\appendix#1#2{\bigskip\noindent%
\centerline{\sectionfont \bf Appendix #1.\ #2}
\bigskip\rm\nobreak}


\newcount\nref
\global\nref = 1

\def\ref#1#2{\xdef #1{[\number\nref]}
\ifnum\nref = 1\global\xdef\therefs{\noindent[\number\nref] #2\ }
\else
\global\xdef\oldrefs{\therefs}
\global\xdef\therefs{\oldrefs\vskip.1in\noindent[\number\nref] #2\ }%
\fi%
\global\advance\nref by 1
}

\def\listrefs{\vfill\eject\section{References}\therefs}


\newcount\cflag
\newcount\nequation
\global\nequation = 1
\def\eqlabel{(1)}

\def\nexteqno{\ifnum\cflag = 0
\global\advance\nequation by 1
\fi
\global\cflag = 0
\xdef\eqlabel{(\number\nequation)}}

\def\lasteqno{\global\advance\nequation by -1
\xdef\eqlabel{(\number\nequation)}}

\def\label#1{\xdef #1{(\number\nequation)}
\ifnum\dflag = 1
{\escapechar = -1
\xdef\draftname{\littlefont\string#1}}
\fi}

\def\clabel#1#2{\xdef\eqlabel{(\number\nequation #2)}
\global\cflag = 1
\xdef #1{\eqlabel}
\ifnum\dflag = 1
{\escapechar = -1
\xdef\draftname{\string#1}}
\fi}

\def\cclabel#1#2{\xdef\eqlabel{#2)}
\global\cflag = 1
\xdef #1{\eqlabel}
\ifnum\dflag = 1
{\escapechar = -1
\xdef\draftname{\string#1}}
\fi}


\def\eeq{}

\def\eqnn #1\eeq{$$ #1 $$}

\def\eq #1\eeq{\xdef\draftname{\ }
$$ #1
\eqno{\eqlabel \rlap{\ \draftname}} $$
\nexteqno}



\def\eol{& \eqlabel \rlap{\ \draftname} \crcr
\nexteqno
\xdef\draftname{\ }}

\def\eeol{& \eqlabel \rlap{\ \draftname}
\nexteqno
\xdef\draftname{\ }}

\def\eolnn{\cr
\global\cflag = 0
\xdef\draftname{\ }}


\def\eqa #1\eeq{\xdef\draftname{\ }
$$ \eqalignno{ #1 } $$
\global\cflag = 0}


\def\ie{{\it i.e.\/}}
\def\eg{{\it e.g.\/}}

\def\etal{{\it et.al.\/}}

\def\myinstitution{
   \centerline{\it Physics Department, McGill University}
   \centerline{\it 3600 University Street, Montr\'eal}
   \centerline{Qu\'ebec, CANADA, H3A 2T8}
}


\def\jetpl#1#2#3#4#5#6{{\it Pis'ma Zh. Eksp. Teor. Fiz.} {\bf #1} (19#2) #3
[{\it JETP Lett.} {\bf #4} (19#5) #6]}

\def\npb#1#2#3{{\it Nucl. Phys.} {\bf B#1} (19#2) #3}
\def\plb#1#2#3{{\it Phys. Lett.} {\bf #1B} (19#2) #3}
\def\prc#1#2#3{{\it Phys. Rev.} {\bf C#1} (19#2) #3}
\def\prd#1#2#3{{\it Phys. Rev.} {\bf D#1} (19#2) #3}

\def\prl#1#2#3{{\it Phys. Rev. Lett.} {\bf #1} (19#2) #3}
\def\rmp#1#2#3{{\it Rev. Mod. Phys.} {\bf #1} (19#2) #3}


\global\nulldelimiterspace = 0pt



\def\frac#1#2{{{#1} \over {#2}}\,}  
\def\hf{{1\over 2}}
\def\nth#1{{1\over #1}}
\def\sfrac#1#2{{\scriptstyle {#1} \over {#2}}}  


\def\Dsl{\hbox{/\kern-.6000em\it D}} 
\def\dsl{\hbox{/\kern-.5600em$\partial$}}
\def\pxpsl{\hbox{/\kern-.5600em$p$}}
\def\ssl{\hbox{/\kern-.5600em$s$}}
\def\epssl{\hbox{/\kern-.5600em$\epsilon$}}
\def\delsl{\hbox{/\kern-.7000em$\nabla$}}
\def\lxpsl{\hbox{/\kern-.5600em$l$}}
\def\kxpsl{\hbox{/\kern-.5600em$k$}}
\def\qxpsl{\hbox{/\kern-.5600em$q$}}
\def\sla#1{\raise.15ex\hbox{$/$}\kern-.57em #1}



\def\roughly#1{\mathrel{\raise.3ex\hbox{$#1$\kern-.75em\lower1ex\hbox{$\sim$}}}}
\def\lsim{\roughly<}
\def\gsim{\roughly>}

\def\ol#1{\overline{#1}}





\def\Scl{{\cal L}}




\def\ket#1{| #1 \rangle}

\def\avg#1{\langle #1 \rangle}



\def\hc{{\rm h.c.}}

\def\sm{standard model}

\def\eV{{\rm \ eV}}
\def\keV{{\rm \ keV}}

\def\GeV{{\rm \ GeV}}

\def\sec{{\rm \ sec}}

\overfullrule=0pt


\ref\kamiokasolar{T.\ Kajita, in the Proceedings of the 25th International
Conference of High Energy Physics, Singapore, K.\ K.\ Phua and
Y.\ Yamaguchi, editors, World Scientific.}

\ref\sage{V.\ N.\ Gavrin, in the Proceedings of the 25th International
Conference of High Energy Physics, Singapore, K.\ K.\ Phua and
Y.\ Yamaguchi, editors, World Scientific.}

\ref\davis{K.\ Lande, in the Proceedings of the 25th International
Conference of High Energy Physics, Singapore, K.\ K.\ Phua and
Y.\ Yamaguchi, editors, World Scientific.}

\ref\bahcall{J.\ N.\ Bahcall \etal, \rmp{54}{767}{1982};
J.\ N.\ Bahcall and R.\ K.\ Ulrich,\rmp{60}{297}{1988}.}

\ref\simpson{J.\ J.\ Simpson, \prl{54}{1891}{1985}.}

\ref\subseq{T.\ Altitzoglou \etal, \prl{55}{799}{1985};
T.\ Ohi \etal, \plb{160}{322}{1985};
V.\ M.\ Datar \etal, {\it Nature} {\bf 318}, 547 (1985);
J.\ Markey and F.\ Boehm, \prc{32}{2215}{1985};
A.\ Apolikov \etal, \jetpl{42}{289}{1985};
D.\ W.\ Hetherington \etal, \prc{36}{1504}{1985}.}

\ref\simparg{J.\ J.\ Simpson, \plb{174}{113}{1986}.}

\ref\recentfor{A.\ Hime and J.\ J.\ Simpson, \prd{39}{1825}{1989};
\prd{39}{1837}{1989};
A.\ Hime and N.\ A.\ Jelley, \plb{257}{441}{1991};
B.\ Sur \etal, \prl{66}{2444}{1991};
I.\ Zlimen, A.\ Ljubicic, S.\ Kaucic, \prl{67}{560}{1991}}

\ref\recentagainst{
H.-W.\ Becker \etal, Caltech preprint 63-605 (1991);
M.\ Bahran, G.R.\ Kalbfleisch Oklahoma preprint OKHEP-91-005, (1991).}

\ref\atmosphere{K.\ S.\ Hirata \etal\rlap, \plb{205}{416}{1988};
D.\ Casper \etal\rlap, \prl{66}{2561}{1991}.}

\ref\Peccei{G.\ Gelmini, S.\ Nussinov, and R.\ D.\ Peccei, UCLA preprint
UCLA/91/TEP/15 (1991), unpublished;
J.\ Cline and T.\ Walker, \prl{68}{270}{1992}.}

\ref\sterilesimpson{D.\ Caldwell and P.\ Langacker, \prd{44}{823}{1991};
E.\ Ma, UC, Riverside preprint UCRHEP-T77, 1991;
K.S.\ Babu, R.N.\ Mohapatra and I.\ Rothstein, \prd{45}{R5}{1991};
A.Yu.\ Smirnov, J.W.F.\ Valle Valencia U. preprint, FTUV-91-38, 1991.}

\ref\GanBurr{
G.\ Raffelt and D.\ Seckel,
\prl{60}{1793}{1988}; K.\ Gaemers, R.\ Gandhi and J.\ M.\ Lattimer,
\prd{41}{2374}{1990};
J.\ Grifols and E.\ Mass\'o, \plb{242}{149}{1990};
A.\ Burrows and R.\ Gandhi, \plb{246}{149}{1990};
erratum: {\it ibid.} {\bf B261}, 519 (1991).}

\ref\SuperTurn{
M.\ Turner, \prd{45}{to appear}{1992}.}

\ref\Seckl{D.\ Seckel and G.\ Raffelt (unpublished).}

\ref\turnerlifetime{
G.\ Steigman and M.\ S.\ Turner, \npb{253}{375}{1985}.}

\ref\efstathiou{J.\ R.\ Bond, G.\ Efstathiou, \plb{265}{245}{1991}.}

\ref\llepton{
M.\ Dugan, G.\ Gelmini, H.\ Georgi and L.\ Hall, \prl{54}{2303}{1985};
O.\ Shanker, \npb{250}{351}{1985};
B.\ Grinstein, J.\ Preskill, and M.\ Wise, \plb{159}{57}{1985}.}

\ref\justso{
S.\ L.\ Glashow, L.\ M.\ Krauss, \plb{190}{199}{1987};
V.\ Barger, R.\ J.\ N.\ Phillips and K.\ Whisnant, \prd{43}{1110}{1991};
A.\ Acker, S.\ Pakvasa and J.\ Pantaleone, {\it ibid.}, 1754.}

\ref\nucleo{R.\ Barbieri and A.\ Dolgov, \npb{349}{743}{1991}.}

\ref\sterileMSW{V.\ Barger, N.\ Deshpande, P.B.\ Pal, R.\ J.\ N.\ Phillips,
and K.\ Whisnant, \prd{43}{1759}{1991};
P.\ Kernan and T.\ Walker, OSU preprint OSU-TA-13/91, 1991.
}

\ref\Barb{R.\ Barbieri and L.\ Hall, \npb{319}{1}{1991}.}

\ref\JerryMarkus{G.\ Jungman and M.\ A.\ Luty, \npb{361}{24}{1991}.}


\def\simp{$17 \keV$ neutrino}

\font\eightrm = cmr8

\def\lblfoot{This work was supported by the Director, Office of Energy
Research, Office of High Energy and Nuclear Physics, Division of High
Energy Physics of the U.S. Department of Energy under Contract
DE-AC03-76SF00098.}

\def\disclaimer{\eject\vskip 3in\littlefont
This document was prepared as an account of work sponsored by the United
States Government.  Neither the United States Government nor any agency
thereof, nor The Regents of the University of California, nor any of their
employees, makes any warranty, express or implied, or assumes any legal
liability or responsibility for the accuracy, completeness, or usefulness
of any information, apparatus, product, or process disclosed, or represents
that its use would not infringe privately owned rights.  Reference herein
to any specific commercial products process, or service by its trade name,
trademark, manufacturer, or otherwise, does not necessarily constitute or
imply its endorsement, recommendation, or favoring by the United States
Government or any agency thereof, or The Regents of the University of
California.  The views and opinions of authors expressed herein do not
necessarily state or reflect those of the United States Government or any
agency thereof of The Regents of the University of California and shall
not be used for advertising or product endorsement purposes.
\vskip 3in\sectionfont
\centerline{\it Lawrence Berkeley Laboratory is an equal opportunity
employer.}
\eject}


\rightline{LBL--31889}
\rightline{McGill--92/03}
\rightline{February 1992}
\title
\centerline{A Natural Framework}
\centerline{for Solar and 17 keV Neutrinos}
\endtitle

\authors
\centerline{C.\ P.\ Burgess and J.\ M.\ Cline}
\vskip .1in
\centerline{\it Department of Physics}
\centerline{\it McGill University}
\centerline{\it 3600 University Street}
\centerline{\it Montr\'eal, Qu\'ebec, Canada H3A 2T8}
\vskip .3in
\centerline{Markus A.\ Luty}
\vskip .1in
\myinstitution
\endauthors

\vfill\eject
\hbox{}\vfil
\abstract
Motivated by recent experimental claims for the existence of a \simp\ and by
the solar neutrino problem, we consider a class of models which contain in
their low-energy spectrum a single light sterile neutrino and one or more
Nambu--Goldstone bosons. In these models, the required pattern of small
neutrino masses and Nambu--Goldstone-boson couplings are understood as the
low-energy residue of the pattern of breaking of lepton-number symmetries near
the electroweak scale, and all mass hierarchies are technically natural. The
models are compatible with all cosmological and astrophysical constraints, and
can solve the solar neutrino problem either via the MSW effect or
vacuum oscillations. The deficit in atmospheric muon neutrinos seen in the
Kamiokande and IMB detectors can also be explained in these models.
\endabstract
\vfil

\disclaimer

\vfil\eject


\section{Introduction}

There are presently several reported experimental anomalies which suggest that
there is new physics lurking in the neutrino sector, and although any one of
these can be incorporated by minimal modifications of the \sm, it is more
difficult to incorporate several of these anomalies simultaneously. It is the
purpose of this paper to argue that the known indiciations of new neutrino
physics can be naturally understood in terms of the low-energy residue of a
particular pattern of lepton-number violation at energies large compared with
the weak scale.

The experimental indications of new neutrino physics are, in what is probably
decreasing order of reliability:

$\bullet$ {\sl The Solar-Neutrino Problem:}
Recent measurements from Kamiokande \kamiokasolar\ and Baksan \sage\ appear to
confirm earlier observations \davis\ of a deficit in the flux of solar
neutrinos as compared to what is predicted by solar models \bahcall. Although
the low event rates make the experiments extremely challenging and
uncertainties linger in the theoretical prediction, the recent confirmations of
the deficit in different types of detectors---including those that are
sensitive to the main $p$--$p$ nuclear cycle in the sun---have given increased
weight to the possibility that new neutrino physics may be responsible. The
most popular proposals for the solution of the solar neutrino problem involve
neutrino oscillations between $\nu_e$ and some other species.

$\bullet$ {\sl The \simp:}
In 1985, Simpson \simpson\ reported experimental evidence for a 17 keV neutrino
which mixes with the electron neutrino at the 10\% level. This claim was very
controversial, since subsequent experiments failed to confirm the effect
\subseq, although Simpson \simparg\ has argued that these experiments were
inconclusive. Recently, there has been renewed interest in the subject, with
several  reports confirming Simpson's results \recentfor, and several others
claiming to rule them out \recentagainst. While it is clear that  the issue of
the existence of the \simp\ is far from settled, it is  striking that the
experimental groups which see the effect  agree very well on the values of the
mass and mixing angle within the experimental uncertainties.

$\bullet$ {\sl The Atmospheric Neutrino Deficit:}
The relative flux of electron- and muon-type neutrinos originating from the
decays of pions produced when cosmic rays impinge on the upper atmosphere
has been measured in several neutrino detectors. These neutrinos are
produced in charged pion decays through the chain $\pi \to \mu \nu_\mu$
followed by $\mu \to e \nu_e \nu_\mu$. The
naive expectation that two $\nu_\mu$'s should be produced for each $\nu_e$
is borne out in detailed simulations which predict $N(\nu_e)/N(\nu_\mu)
= 0.45$. This ratio as measured by both Kamiokande and IMB \atmosphere\ is
larger than predicted.  These observations could be accounted for by
near-maximal mixing of $\nu_\mu$ with another species of neutrino.

None of these experimental results is beyond controversy at present,  although
the solar-neutrino results are probably the least controversial. More
experiments are currently  underway to determine which (if any) of these
effects are real.

Taken separately, each of the neutrino results can be easily accounted for in
terms of a particular form for the masses and mixing of the three known
neutrino types. However, we wish to argue that the solar neutrino problem and
the existence of the \simp, together with current cosmological and
astrophysical bounds, point toward a specific form for the neutrino mass matrix
which can arise naturally from new physics at high scales. This same form for
the mass matrix can also account for the atmospheric muon-neutrino deficit.
While we feel that it is certainly premature to take {\it all} of these results
seriously, we feel that it is still interesting to see that they can all be
accomodated in a rather simple and natural framework. We will therefore suspend
our disbelief, and address ourselves to the question of how the neutrino masses
and mixing needed to solve the solar neutrino problem and incorporate the
\simp\ can be added to the standard model.

This paper is organized as follows:
In section 2, we briefly recount the constraints on the properties of a \simp.
We argue that the existence of a \simp, together with a neutrino solution
to the solar neutrino problem, requires the existence of a light sterile
neutrino species, and suggests the existence of Nambu--Goldstone
bosons (majorons).
In sections 3 and 4, we derive the general form of the interactions of the
neutrinos and majorons at energies below the weak scale, and discuss how these
are constrained by laboratory experiments and cosmological and astrophysical
arguments.
In sections 5 through 7, we explore the implications of this phenomenology
at higher energies.
Section 5 gives a statement of the naturalness requirements to which we adhere
in our exploration of the candidate models for high-energy physics
that might produce the `observed' neutrino spectrum.
Sections 6 and 7 give examples of models which satisfy these criteria.
Our conclusions are summarized in section 8.

\section{Implications of a 17 keV Neutrino}

There are many constraints on the properties of the \simp, which are usefully
reviewed \eg\ in \Peccei. It cannot be mainly the muon neutrino, since direct
bounds on $\nu_e$--$\nu_\mu$ oscillations already rule out a $10\%$ mixing. In
order to avoid conflicting with double beta decay experiments, the contribution
of the \simp\ to the rate for neutrinoless double beta decay must be accurately
cancelled by the contributions from other neutrino states. This cancellation
arises most naturally if there are two neutrino states of opposite $CP$ parity
with Majorana masses close to 17 keV; that is, if the \simp\ is a Dirac (or
pseudo-Dirac) state. In this case the suppression of the neutrinoless
beta-decay rate can be understood as being due to the approximate conservation
of a quantum number carried by the \simp. It will turn out that this quantum
number can be only approximately conserved if the solar neutrino problem is
solved by neutrino oscillations.

This type of neutrino mass spectrum may be obtained using only the  three known
neutrinos if the $\nu_\mu$ and $\nu_\tau$ form the  nearly degenerate \simp. If
this is the case, then there is no way to solve the solar neutrino  problem by
neutrino mixing, since such a solution would require  another neutrino state
with mass less than 10 eV which can  mix with $\nu_e$. The solar neutrino
problem and the \simp\ taken together therefore require the existence of at
least one new neutrino species, $s$, beyond the three already observed. This
new state must be sterile, \ie\ it cannot carry $SU(2)_W \times U(1)_Y$ quantum
numbers, since it was not observed in the $Z$ width at LEP.

There are therefore two natural possibilities: either the sterile state forms
part of the \simp, or it mixes  with the electron neutrino to solve the solar
neutrino problem. Suppose that the first possibility holds
\sterilesimpson. In this case there is a stringent bound coming
from the energetics  and timing of the observed neutrino pulse from the
supernova SN1987A.
The idea is that helicity-flipping processes can produce the  sterile
state in the core, resulting in rapid core cooling  via emission of sterile
neutrinos. Early work \GanBurr\ on this subject  gave a bound of $m_D \le 28
\keV$ (when corrected for an  erroneous factor of 4), but there have been
subsequent claims  \SuperTurn\ that effects such as neutrino degeneracy will
lower the bound to $\simeq 1 \keV$. The situation is not yet  settled, since
there are other competing contributions which have  not yet been included in
any detailed numerical calculation \Seckl.
Despite the uncertainties in the
supernova bound, we will not pursue this possibility here, and concentrate
instead on the scenario in which the electron neutrino mixes with the sterile
neutrino state \sterilesimpson\ to solve the solar neutrino problem, and
$\nu_\mu$ and $\nu_\tau$ pair into a pseudo-Dirac \simp\ state that mixes with
$\nu_e$  at the 10\% level.

Cosmological constraints on massive neutrinos suggest that the
low-energy spectrum of the theory must be enlarged even further. If a neutrino
species with mass in the range $100 \eV \roughly< m_\nu \roughly< 1 \GeV$ were
absolutely stable and in chemical equilibrium, its present energy density would
dominate the universe, resulting in an unacceptably young universe. A mechanism
is therefore required to deplete the number density of the \simp.

It is possible that the \simp\ is stable and that its number density in the
early universe is depleted by annihilation mediated by some new interactions,
but it is far more natural simply to make the \simp\ unstable. (Standard
arguments to this effect are reviewed \eg\ in \Peccei.)  The lifetime that is
required is shorter than $\sim 10^{12} \sec$.
The only standard-model candidates for the decay products of a \simp\ are
$\nu_{17} \to \nu \gamma$ or $\nu_{17} \to 3\nu$.
The decay into photons is severely constrained and exotic interactions are
required to make the three-neutrino decay mode sufficiently rapid.
However, we find it simpler to posit another light particle into which the
\simp\ can decay.

In fact, a candidate for such a light particle arises naturally in the class
of models we will be considering.
In these models, the approximate symmetries that suppress neutrinoless
double-beta decay are assumed to be broken {\it spontaneously}.
The fact that these symmetries are still approximate is explained by the fact
that the symmetry breaking sector is weakly coupled to observed particles as
an automatic consequence of the quantum numbers of the order parameter.
In this case, the theory automatically contains massless Nambu--Goldstone
bosons (majorons) which are weakly coupled to the neutrinos, and allow the
decay mode $\nu_{17} \to \nu' \chi$, where $\chi$ is a majoron.

This mechanism is not the only way to incorporate such majorons.
An alternative would be to consider Nambu--Goldstone bosons arising from the
spontaneous breaking of a larger symmetry group which contains the approximate
symmetries in our models.
Or the broken symmetries could be both explicitly and spontaneously broken in
the underlying theory'
In this case, the majoron would be a {\it pseudo\/}-Nambu--Goldstone boson
with a mass and non-derivative interactions whose sizes are determined by the
strength of the explicit breaking of the symmetry.
However, we concentrate on the first option because it is
more constrained, and because it connects the existence of the majorons
directly with the origin of the approximate symmetries which are anyhow already
required at low energies.

There is an additional cosmological constraint on neutrino lifetimes which can
be derived from considerations of structure formation. In the standard
scenario, the structure observed in the universe today is formed by the
gravitational amplification of small density perturbations in the early
universe. This amplification cannot occur during a radiation-dominated epoch,
and demanding that the decay products of the \simp\ do not overly prolong this
epoch gives a lower bound on its lifetime.
According to  ref.\ \turnerlifetime\ the standard scenario remains undisturbed
provided that the lifetime is shorter than $\sim 10^6$ sec.
However, some recent studies of large scale structure \efstathiou\ indicate
that if the 17 keV neutrino lifetime were as large as $10^7$--$10^8$ secs it
might actually improve the status of cold dark matter models by enhancing the
strength of correlations of density perturbations at long distances.
We will find that there are models which satisfy all other bounds but which
are in conflict with the structure formation bounds.
However, the paradigm for structure formation is not well-tested, and we
therefore do not consider such models to be ruled out.

A final constraint arises if the final state for \simp\ decay should include
$\ol{\nu}_e$'s. If so then the light products of heavy neutrinos that decay
in flight while en route from SN1987a can arrive much later than those that
are emitted directly from the core.
Agreement between the length of the observed pulse and supernova models then
requires that the lifetime not lie between $3 \times 10^4$ and
$2 \times 10^8$ secs.

\section{Neutrino Masses and Mixings}

Beta-decay experiments, solar neutrino measurements, and atmospheric neutrinos
all probe neutrino properties at energies very low compared to the weak scale.
Their implications for the neutrino sector may therefore be most succinctly
expressed in terms of the properties of the low-energy theory obtained after
integrating out all particles that are heavier than 17 keV. In this section we
collect the implications for this low-energy theory of the recent neutrino
results. These are used in subsequent sections to infer some of the properties
of the underlying physics at higher energies that might be responsible for such
an effective theory.

In the standard model the spectrum at extremely low energies contains four
exactly massless particles: three left-handed neutrino flavours $\nu_e$,
$\nu_\mu$, and $\nu_\tau$, and the photon. The masslessness of the neutrinos
can be explained by the conservation of the three lepton numbers, while the
masslessness of the photon is explained by electromagnetic gauge invariance.

Motivated by the arguments of the previous section, we suppose that this
spectrum is supplemented by at least two additional states---a single sterile
fermion, $s$, and a light pseudo-Nambu-Goldstone boson, $\chi$. Like the
photon, $\chi$ is kept massless (and, at the renormalizable level,
noninteracting) by a symmetry: $\chi \to \chi + f$ with  $f$ an arbitrary
constant.

As for the conserved lepton numbers, all three cannot be symmetries of  the
low-energy lagrangian if it is to naturally account for the \simp\  and to
solve the solar neutrino problem, since the \simp\ must be unstable, and
$\nu_e$ must oscillate into another light state. Instead, a symmetry is
required that can ensure that the \simp\ is a $\nu_\mu$--$\nu_\tau$
pseudo-Dirac state and which allows this state to mix with $\nu_e$ at the
$10\%$ level. The symmetry must also ensure that the sterile neutrino remains
sufficiently light that its mixing with $\nu_e$ can deplete the observed solar
neutrino flux.

The pseudo-Dirac nature of the 17 keV state and its mixing with $\nu_e$ is
ensured if the theory approximately preserves the linear combination  $e - \mu
+ \tau$ of the standard model lepton numbers \llepton. The absence of a large
majorana mass for the sterile fermion, $s$, suggests a further approximate
$U(1)$ chiral symmetry which may be defined so that it rephases only $s$. We
therefore assume that the low-energy lagrangian approximately preserves the
symmetry
\eq
G_\nu \equiv U(1)_{e-\mu+\tau} \times U(1)_s
\eeq
under which the left-handed neutrino fields transform as
\eq
\nu_e, \nu_\tau \sim (1,0), \qquad
\nu_\mu \sim (-1,0), \qquad
s \sim (0,1).
\eeq
Of course, $G_\nu$ cannot be an exact symmetry, since it also forbids the
$\nu_e$--$s$ oscillations that are to account for the solar-neutrino
deficit. $G_\nu$ must therefore be only an {\it approximate} symmetry
of the low-energy theory. More will be said about the origins of this symmetry
breaking once we discuss explicit models for the underlying physics.

Subject to these assumptions the neutrino mass terms must take the following
form when expressed in terms of a weak-interaction basis of left-handed
fields:
\eq
\label\basis
\Scl_m = -\hf \, \pmatrix{s \cr \nu_e \cr \nu_\mu \cr \nu_\tau \cr}^T
\Bigl( M_0 + \delta M \Bigr)
\pmatrix{s \cr \nu_e \cr \nu_\mu \cr \nu_\tau \cr} + \hc
\eeq
Here $M_0$ is $G_\nu$-invariant but $\delta M \ll M_0$ is not.
We write
\eq\label\massmatrix
M_0  = m_{17} \pmatrix{0&0&0&0\cr 0&0&s&0\cr 0&s&0&c\cr 0&0&c&0\cr}, 
\qquad\delta M  = m_{17} \pmatrix{\gamma & \alpha_1 & \beta & \alpha_2 \cr
\alpha_1 & \epsilon_1 & 0 & \epsilon_2 \cr
\beta & 0 & \eta & 0 \cr
\alpha_2 & \epsilon_2 & 0 & \epsilon_3 \cr}, 
\eeq
where $s=\sin\theta_{17}$, $c=\cos\theta_{17}$ and $\theta_{17}$ is the
$\nu_e$-$\nu_{\tau}$ mixing angle. For simplicity we choose the elements of
$\delta M$ to be real. The notation is chosen such that matrix
elements that are represented by the same Greek letters transform identically
with respect to the symmetry group $G_\nu$.
Since mass-matrix elements that transform in the same way should be of the
same order of magnitude, this notation is useful when choosing the
symmetry-breaking patterns that are
required to produce the `observed' heirarchies in the mass matrix.

It is often convenient to refer to the rotated basis
\eq
\eqalign{
\ket{\nu_{e}'} & = c\ket{\nu_e}-s\ket{\nu_{\tau}}, \cr
\ket{\nu_{\tau}'} & = c\ket{\nu_{\tau}}+s\ket{\nu_e}, \cr}
\eeq
in which
\eq
M_0' = m_{17} \pmatrix{0&0&0&0\cr 0&0&0&0\cr 0&0&0&1\cr 0&0&1&0\cr}, 
\qquad\delta M' = m_{17} \pmatrix{\gamma & \alpha_1' & \beta & \alpha_2' \cr
\alpha_1' & \epsilon_1' & 0 & \epsilon_2' \cr
\beta & 0 & \eta & 0 \cr
\alpha_2' & \epsilon_2' & 0 & \epsilon_3' \cr}. 
\eeq
(The only relations between the primed and unprimed matrix elements which will
be needed in the following are $\alpha_1' = c \alpha_1 - s \alpha_2$ and
$\alpha_2' = c \alpha_2 + s \alpha_1$.)

In what follows, we will assume that the low-energy theory breaks $G_\nu$ via
order parameters transforming under $G_\nu$ in specified ways. The choice of
order parameters will determine the hierarchy of the elements of $\delta M$. In
order to determine the order parameters required, we first discuss the
phenomenology which results from the mass matrix of eq.\ \basis.

In the limit $\delta M \to 0$, the spectrum consists of a massive Dirac state
\eq
\ket{\nu_{h\pm}} = \frac 1{\sqrt{2}}
\left(\ket{\nu_\tau'} \pm \ket{\nu_\mu} \right)
\eeq
with mass $m_{h\pm} = m_{17}$, together with two massless states. We can
compute
the spectrum for $\delta M \ll M_0$ using standard degenerate perturbation
theory. To second order in $\delta M$, the heavy states become split with
\eqa
\Delta m_h^2 & \equiv m_{h+}^2 - m_{h-}^2 
 = 2{m_{17}^2}\left[\epsilon_3'+ \eta
  +2\beta \alpha_2' \right] + O((\delta M)^3). 
\eeq
To first order in $\delta M$, the massless states acquire masses
\eq
m_{\ell\pm} = \frac {m_{17}}2 \; \left( \lambda \pm \Delta \right),
\eeq
where
\eqa
\Delta & \equiv
\gamma + \epsilon_1', \eol
\lambda & \equiv \sqrt{\left( \gamma - \epsilon_1' \right)^2
+ 4 \alpha_1'^2}^{1/2}. \eeol
\eeq
The mass splitting of the light states is
\eq
\Delta m_\ell^2 = m_{17}^2 \lambda \Delta.
\eeq
To first order in $\delta M$, the light eigenstates are given by
\eq\eqalign{
\ket{\nu_{\ell+}} & = \cos\theta_\ell \ket{s}
+ \sin\theta_\ell \ket{\nu_e'}, \cr
\ket{\nu_{\ell-}} & = \cos\theta_\ell \ket{\nu_e'}
- \sin\theta_\ell \ket{s}, \cr}
\eeq
where
\eq
\tan2\theta_\ell = \frac{2\alpha_1'}{\gamma-\epsilon_1'}.
\eeq
We now have in hand the physical quantities that arise in the neutrino
phenomenology in terms of the properties of the neutrino mass matrix.
The constraints on the parameters introduced above are as follows:

$\bullet$ {\sl Laboratory mass bound:}
The present bound $m_{\nu_e} < 9 \eV$  on the mass of the electron
neutrino implies a similar bound on the mass of the light neutrino state that
dominantly overlaps $\nu_e$. In the models we will consider, this constraint is
easily satisfied.

$\bullet$ {\sl The 17 keV neutrino:}
The experiments which see a \simp\ find that it is produced in approximately
1\% of beta decays. This requires
\eq\eqalign{
m_{17} & = 17 \keV, \cr
\sin\theta_{17} & \simeq 0.1. \cr}
\eeq

$\bullet$ {\sl $\nu_\mu$--$\nu_\tau$ oscillations:}
Because $\nu_\mu$--$\nu_\tau$ mixing is nearly maximal, the failure to observe
$\nu_\mu$ disappearance at Fr\'ejus implies
$\Delta m_h^2 \le 5 \times 10^{-3} \eV^2$, which gives
\eq
\delta_h \equiv \epsilon_3' + \eta
+ 2\beta \alpha_2'\le 2 \times 10^{-11}.
\eeq

$\bullet$ {\sl Atmospheric neutrinos:}
The atmospheric neutrino anomaly reported by Kamiokande and IMB can be
explained by near-maximal $\nu_\mu$--$\nu_\tau$ mixing provided that
$\Delta m_h^2 \ge 5 \times 10^{-4} \eV^2$, which gives
\eq
\delta_h \gsim 2 \times 10^{-12}.
\eeq

$\bullet$ {\sl The solar neutrino problem:}
$\nu_e$--$s$ oscillations may deplete the solar neutrino flux observed on earth
either through resonant MSW oscillations in the sun or through maximal vacuum
oscillations. Resonant oscillations are the currently favored mode of solution
given the small size of the flux measured by the Chlorine experiment. Maximal
vacuum oscillations tend to reduce the solar neutrino flux by an overall factor
of two, unless the oscillation length happens to be close to the earth-sun
distance---so-called ``just-so'' oscillations \justso. A factor-of-two
suppression would be in agreement with the Kamiokande measurement, but well
outside of the 90\% confidence level upper bound for the Chlorine experiment if
we use the theoretical predictions of ref.\ \bahcall. Nonetheless, in what
follows, we will entertain the idea that maximal mixing with an overall
neutrino flux suppression of one half may turn out to be the correct solution
of the solar neutrino problem, and we will consider this scenario alongside the
more traditional MSW and ``just-so'' scenarios. The reader is free to disregard
this region of parameter space.

Although we are working with a four-state system, it is clear that the
\simp\ is too massive to be relevant for the solar neutrino problem.
Therefore, we can reduce the problem to that of mixing between the states
$\ket s$ and $\ket{\nu_e'}$.
The parameter regions that are allowed for the different solutions to the
solar neutrino problem are as follows:

$\bullet$ {\sl Maximal vacuum oscillations:}
Maximal vacuum oscillations can ``solve'' the solar neutrino problem provided
that $\sin^2 2\theta_\ell \simeq 1$ and $\Delta m_\ell^2 \gsim 10^{-10} \eV^2$.
The lower limit of this mass range corresponds to ``just-so'' oscillations
\justso. In addition, there is a cosmological bound arising from the
observation that maximal $\nu_e$--$s$ oscillations can change the number
density of $\nu_e$'s required for the standard model of big-bang
nucleosynthesis. This bound is $\Delta m_\ell^2 \leq 2\times 10^{-7} \eV^2$.
Putting this together, we find the restrictions
\eq
3 \times 10^{-19} \lsim \lambda\Delta \lsim 6 \times 10^{-16},
\eeq
and
\eq
\gamma - \epsilon_1'\ll
2 \alpha_1'.
\eeq

$\bullet$ {\sl Resonant oscillations:}
Resonant MSW oscillations require
$10^{-4} \lsim \sin^2 2\theta_\ell \lsim 0.7$, which gives
\eq
1 \lsim \frac{\gamma - \epsilon_1'}{
\alpha_1'} \lsim 200.
\eeq
Because Kamiokande II observes some solar $\nu_\mu$'s, comparison with SAGE
and the $^{37}$Cl date can be used to distinguish between
$\nu_e\leftrightarrow\nu_\mu$ and $\nu_e\leftrightarrow s$ oscillations.  This
has been studied in detail \sterileMSW\ with the result that only the
nonadiabatic branch of the MSW triangle is allowed, and it is shifted to
somewhat lower values of $\Delta m^2$ relative to the ordinary MSW effect.
Numerically, this branch is specified by
\eq
\Delta m_\ell^2 \; \frac{\sin^2 2\theta_\ell}{\cos 2\theta_\ell}
\simeq 6.8 \times 10^{-8} \eV^2,
\eeq
which gives
\eq
\alpha_1' \simeq 8\times 10^{-9}
\eeq

\section{Majoron Couplings}

The low-energy interactions of Nambu--Goldstone bosons are largely dictated
by the symmetry-breaking pattern which give rise to them.
This allows us to treat the majorons which are assumed to appear in our
models in our general framework.
The lowest-dimension interactions between the neutrinos and majorons have
dimension five:
\eq
\label\gbcoupling
\Scl_\chi = - \nth{f} \; \partial_\mu \chi \; J^\mu,
\eeq
where $J^\mu$ is the conserved current which is spontaneously broken at the
scale $f$.   (If the symmetry is broken by a set of fields $\Phi_a$ whose
charges and VEV's are $q_a$ and $v_a$ respectively, then $f=2(2\sum_a q^2_a
v^2_a)^{1/2}$).
The coupling to neutrino species $\nu_j$ ($j = s, e, \mu, \tau$) is therefore
determined by its quantum numbers with respect to the broken symmetry.
We write
\eq
J^\mu = i \ol{\nu}_j \gamma^\mu Q_{jk} \gamma_5 \nu_k,
\eeq
where $Q_{jk}$ is the hermitian matrix which generates the symmetry on a basis
of left-handed fields.
If the left-handed fermions are rotated to a mass eigenbasis via a unitary
matrix $U$, then the corresponding charge in terms of the mass basis becomes
$Q' = U^\dagger Q U$.

As discussed in previous sections, the most economical assumption is that
broken symmetry to which the majorons couple is $G_\nu$ itself.
In this case the generators that represent the two factors of this symmetry on
the left-handed neutrino fields are both diagonal in the weak-interaction
basis:
\eq
S = \pmatrix{1 &&&\cr &0&&\cr &&0&\cr &&&0 \cr} \quad \hbox{and}
\quad
L = \pmatrix{0 &&&\cr &1&&\cr &&-1&\cr &&&1 \cr}.
\eeq
In the mass basis, these charges become
\eqa
S'_{ab} &= U^*_{sa} U_{sb} \eolnn
L'_{ab} &=  U^*_{ea} U_{eb} - U^*_{\mu a} U_{\mu b} + U^*_{\tau a}
U_{\tau b}. \eeol
\eeq
Here $a,b = \ell\pm, h\pm$ label the mass eigenstates.

\def\sss{\scriptscriptstyle}
\def\cl{c_{\sss L}} \def\sl{s_{\sss L}} \def\ap{\alpha_2'}
\def\ep{\epsilon_2'} \def\bp{\beta}

For both of these generators, the matrix elements that link the heavy with
light states vanish at zeroeth order in $\delta M$.  The leading contributions
are most conveniently tabulated for the linear combinations $L'\pm S'$,
\eq
\label\Smaj
\eqalign{
(L'+S')_{l\pm,h\pm} &=\sqrt{2}\pmatrix{\cl\ap+\sl\ep & \cl\ap+\sl\ep \cr
				     -\sl\ap+\cl\ep &-\sl\ap+\cl\ep \cr};
\cr
(L'-S')_{l\pm,h\pm}& =\sqrt{2}\pmatrix{\cl\bp+\sl\ep &-\cl\bp+\sl\ep \cr
				     -\sl\bp+\cl\ep & \sl\bp+\cl\ep \cr},
\cr}
\eeq
where $\cl=\cos\theta_\ell$ and $\sl=\sin\theta_\ell$.  For a given choice of
symmetry-breaking scalar fields, one can find two orthogonal directions in
field space corresponding to the two Majorons.  Each couples to its own
particular linear combination $Q'$ of the above charges.  Then the partial
lifetime for the decay of $\nu_h$ into that Majoron and one of the two light
neutrino states is
\eqa
\tau(\nu_h \to \nu_\ell \chi) & =
\frac{4\pi f^2}{m^3_h |Q'_{h\ell}|^2 } \eolnn
\label\nulife
& = \left(1.7 \times 10^{-5} \sec \right)
\left( \frac f{100 \GeV} \right) ^2 \frac 1{|Q'_{h\ell}|^2}. \eeol
\eeq
Satisfaction of the cosmological bound coming from the age of the universe
therefore requires $f/|Q'_{h\ell}|^2 \lsim 2\times 10^{10} \GeV$ and the
structure-formation bound is 100 times smaller.

\section{Naturalness Criteria}

We wish to show that the desired neutrino mass pattern can arise in ``natural''
models. Our criteria for naturalness are supposed to capture the idea that we
do not want to give up any of the successes of the standard model, and we do
not want to add to its shortcomings.\footnote{$^\dagger$}{This succinct
formulation is taken from R.\ Barbieri and L.\ Hall, ref.\ \Barb.}
Specifically, our requirements are as follows:

$\bullet$ We demand that the model have no new symmetry breaking scales below
the weak scale. The reason for this condition is that all of  the known ways of
understanding the magnitude of the weak  scale (compared, say, to the Planck
scale) necessarily involve  new particles and interactions not far above the
weak scale.  If we were to introduce a new symmetry breaking scale below the
weak scale, it would be difficult to imagine how such a heirarchy  could be
explained without introducing new light particles  which should  already have
been observed.

In fact, in the context of the models we discuss below, the smallness of the
neutrino masses compared to the electroweak scale is due to physics at a very
large scale, $M \gg M_W$. It is therefore useful to view the standard  model as
an effective theory which is valid below the scale $M$. We can summarize the
low-energy effects of the physics above the scale $M$ by including all possible
higher-dimension operators in the effective lagrangian. The coefficients of the
higher-dimension operators in this lagrangian are proportional to inverse
powers of $M$, and so their effects are typically suppressed at low energies.

$\bullet$ We also demand that the magnitudes of all small parameters be
understood in terms of symmetry principles.
This criterion comes in two parts:
First, supposing that a parameter, such as a neutrino mass, should turn out
to be small in the underlying microphysical lagrangian above the scale $M$,
we require that the smallness of this parameter should be stable under its
renormalization to lower energy scales where it is measured.
This is ensured if the small parameter satisfies the naturalness criterion of
't Hooft, according to which a parameter is naturally small if additional
symmetry arises in the limit that the parameter in question vanishes.
To the extent that the
renormalization process preserves this symmetry the vanishing of the
symmetry-breaking parameter must be stable under the renormalization and any
deviations from zero that are generated by renormalization are automatically
proportional to the original value of the parameter itself. The electron mass
is a familiar example of a parameter that is naturally small according to this
criterion, since the standard model acquires an extra chiral symmetry in the
limit that the electron mass vanishes.

Of course, naturalness in this technical sense does not address the question of
why the parameter is small in the underlying theory in the first place. The
second part of our criterion follows from the motivation that we would
ultimately like some understanding of the origin of the smallness of a
parameter in the underlying high-energy theory. When we turn to models of
physics at the scale $M$ we therefore propose that small parameters such as
neutrino masses can be understood in terms of a hierarchy of symmetry-breaking
scales, which are themselves protected by a symmetry. An attractive feature of
the specific models we will discuss is that they require the introduction of
only one new large scale $M$, and all small parameters are understood in terms
of the hierarchy $v / M$.

$\bullet$ We assume that the only light degrees of freedom that appear in the
effective theory at and below electroweak scales are the usual standard model
particles (including a single Higgs doublet), supplemented by the minimal
number of additional degrees of freedom that are required to account for
the solar-neutrino problem and the \simp.
As discussed above, we take these to be a single electroweak-singlet fermion
and (at least) one electroweak singlet Goldstone boson into which the
\simp\ can decay.
We do not address the hierarchy problem associated with the standard
Higgs field, since we do not expect this to be more difficult to solve
here than within the standard model, using supersymmetry,
for example.

We next turn to the construction of explicit models which
produce the desired low-energy behaviour in a natural way.

\section{A Vacuum-Oscillation Model}

The pattern of neutrino masses in our framework is determined by the
hierarchies in the mass matrix $\delta M$. This, in turn, is predominantly
controlled by the quantum numbers of the order parameters that break $G_\nu$.
As might be expected, the required quantum numbers differ significantly
depending on whether the solar neutrino problem is solved through resonant or
maximal vacuum oscillations, and we treat these cases separately. In this
section, we present a model with maximal vacuum oscillations.

In order to systematically build in our naturalness requirements, we begin our
analysis at the level of an effective theory valid at the scale at which the
symmetry $G_\nu$ breaks. This scale will turn out to be near the weak scale.

\noindent$\bullet$ {\it Weak-Scale Effective Theory}

The degrees of freedom at the scale at which $G_\nu$ is broken are assumed to
be the usual standard-model fields, together with the gauge-singlet fermion $s$
and two gauge-singlet scalar fields $\phi_1$ and $\phi_2$ transforming under
$G_\nu$ as
\eq
\phi_1 \sim (\sfrac 12,-\sfrac 12) \qquad\hbox{and}\qquad
\phi_2 \sim (-\sfrac 12, -\sfrac 12).
\eeq
The lowest-dimension gauge- and $G_\nu$-invariant operators in the effective
lagrangian at this scale that contribute to the neutrino mass matrix are
\eq
\label\themodelI
\eqalign{
\hbox{dimension 5:} \quad & \frac{g_e}{M} \, (L_e H)(L_\mu H),
\quad \frac{g_\tau}{M} \, (L_\mu H)(L_\tau H); \cr
\hbox{dimension 6:} \quad & \frac{a_j}{M^2} \, (L_j H) \, s \phi_2^2,
\quad \frac{b}{M^2} \, (L_\mu H) \, s \phi_1^2;\cr
\hbox{dimension 7:} \quad & \frac{c}{M^3} \, ss \phi_1^2 \phi_2^2; \cr
\hbox{dimension 9:} \quad & \frac{d_{\mu\mu}}{M^5} \,
(L_\mu H)(L_\mu H) (\phi_1 \phi_2^*)^2,
\quad \frac{d_{jk}}{M^5} \, (L_j H)(L_k H) (\phi_1^* \phi_2)^2. \cr}
\eeq
Here $H$ is the usual electroweak Higgs doublet, the $L$'s  are the standard
left-handed lepton doublets, and $j, k = e, \tau$ are generation indices.
Explicit factors of a heavy mass scale $M$ have been included so that the
coefficients of these operators in the effective lagrangian are dimensionless.
If these operators arise from new physics at the scale $M$, and there are no
symmetries beyond those we have assumed, then all of the coefficients of these
operators will be of order unity in the absence of fine-tuning.

If we replace the scalars with their vacuum expectation values
\eq
\avg{H} = v = 174 \GeV, \qquad
\avg{\phi_1} = w_1, \qquad
\avg{\phi_2} = w_2,
\eeq
and define $g=\sqrt{g_e^2+g_\tau^2}$, then the mass-matrix parameters of
eq.\ \massmatrix\ are
\eq
\eqalign{
M & = 1 \times 10^7 g v \simeq 2 \times 10^9 g \GeV, \cr
\alpha_j & =  \frac{a_j w_2^2}{g Mv}, \cr
\beta & = \frac{b w_1^2}{g Mv}, \cr
\gamma & = \frac{c w_1^2 w_2^2}{g M^2 v^2}, \cr
\epsilon_j, \eta & = \frac{d w_1^2 w_2^2}{g M^4}.
\cr}
\eeq
Assuming that $v, w_1, w_2 \ll M$, one has the hierarchy
$\epsilon,\eta \ll \gamma \ll \alpha, \beta$.
In this case, the light neutrino states form a pseudo-Dirac
pair with
\eqa
m_\ell & = m_{17} \alpha_1', \eol
\Delta m_\ell^2 & = m_{17}^2 \gamma \alpha_1', \eeol
\eeq
while the heavy neutrino states have mass splitting
\eq
\Delta m_h^2 = 4m_{17}^2 \beta \alpha_2'.
\eeq

There are two majorons in this model, $\chi_1$ and $\chi_2$, which can be
thought of as the phases of the fields $\phi_1$ and $\phi_2$, respectively.
$\chi_1$ couples to the charge $Q_1 \equiv S - L$, while $\chi_2$ couples to
$Q_2 \equiv S + L$. The decay constants are related to the corresponding vacuum
expectation values by $f = 2\sqrt{2} w$.  Using eqs.\ \Smaj\ and \nulife, the
lifetime is
\eq
\tau = \frac{16\pi}{m_h^3} \left( \frac{\alpha_2'^2}{w_2^2} +
\frac{\beta^2}{w_1^2} \right)^{-1}
\eeq
In section 2 it was noted that in order to satisfy constraints from SN1987a
and cosmology, the lifetime of the \simp\ should either be $< 10^4$ sec or
$\sim 10^9$ sec.
Either possibility can be accomodated in this model.

For the case of long lifetimes, we find that all the phenomenological
constraints can be satisfied by choosing
$w_1, w_2 \sim 3v$, $a_j = b = c = g_\tau = 1$ and $g_e = 0.1$.  Then
$\alpha, \beta \sim 10^{-6}$, $\gamma \sim 10^{-12}$, and the lifetime is
$10^9$ sec.  The neutrino masses are given by
\eq
\label\splittings
\eqalign{
m_\ell & \sim 0.01 \eV, \cr
\Delta m_\ell^2 & \sim 10^{-10} \eV^2, \cr
\Delta m_h^2 & \sim 10^{-3} \eV^2. \cr}
\eeq
Note that the hierarchies $m_h / v$ and $m_\ell / m_h$, as well as  $\Delta
m_h^2 / m_h^2$ and $\Delta m_\ell^2 / m_\ell^2$ have been explained by the
largeness of $M$ relative to $v$, $w_1$ and $w_2$.
It is interesting that both
$w_1$ and $w_2$ preferentially lie near the weak scale because of the
$\nu_\mu$--$\nu_\tau$ oscillation bound and our naturalness condition that
there be no symmetry-breaking scales below the weak scale. $\Delta m_h^2$ is
then near the experimental upper limit and in the range required to account for
the atmospheric neutrino anomaly. Also, $\Delta m^2_\ell$ falls naturally into
the correct range for ``just-so'' vacuum oscillations.

Although not a generic prediction of this model, fast \simp\ decays can be
obtained by taking $w_1 = 100v$, $w_2=v/3$, $a_1=0.1$, $a_2 =0.01$, $b=1$,
and $c=g= 0.3$.  Then $\alpha_1 = 10^{-8}$, $\alpha_2 = 10^{-9}$, $\beta =
10^{-2}$, $\gamma = 10^{-10}$ and the lifetime is $2\times 10^3$ sec.  The mass
splittings are the same as in \splittings, but $m_\ell$ itself is now only
$10^{-4}$ eV.

\noindent$\bullet$ {\it Renormalizable Model}

Here we present a renormalizable model defined at scale $M$ which can give rise
to the weak-scale lagrangian just described. This is done only as an existence
proof, since there are clearly many possible models, and it is unlikely that
any forseeable experiment could distinguish among them. In constructing a
renormalizable model, we are guided solely by principles of economy.

The model contains the fields previously described with the addition of four
gauge-singlet Dirac fermions.  In terms of left-handed fields, they transform
under $G_\nu$ as
\eq
N_{1}^{\pm} \sim \pm(\sfrac 12, \sfrac 12),
\qquad N_{2}^{\pm}  \sim \pm(\sfrac 12, -\sfrac 12),
\qquad\ N_{3,4}^{\pm}  \sim \pm(1, 0),
\eeq
Two copies of the last charge assignment are required in order to avoid an
accidental symmetry of the neutrino mass matrix which forces two of the light
states to be massless.

The renormalizable interactions of this model are the usual standard model
interactions, with the addition of
\eq\eqalign{
\hbox{dimension 3:} \qquad & M_j N_j^+ N_j^-,\cr
\hbox{dimension 4:} \qquad & (L_\mu H) N^+_{3,4},
\quad (L_j H) N_{3,4}^-, \quad s N_1^- \phi_1, \quad s N_2^+ \phi_2, \cr
& N^-_{3,4} N_1^+ \phi_1, \quad N_{3,4}^+ N_1^- \phi_1^*, \quad
N_{3,4}^- N_2^+ \phi_2^*, \quad N_{3,4}^+ N_2^- \phi_2. \cr}
\eeq
If we assume that all dimensionless coefficients are of order $0.1 - 1$,
then the heavy fermions have masses of order $10^{8} \GeV$.
It is easy to check that when the heavy fermions are integrated out, the
resulting weak-scale effective theory is exactly the one described above.

\section{An MSW Model}

We now turn to the construction of a model which can solve the solar neutrino
problem via resonant MSW oscillations.  As in the previous section, we will
find that $G_\nu$ is preferentially broken near the weak scale.

\noindent$\bullet$ {\it Weak-Scale Effective Theory}

The degrees of freedom at the $G_\nu$-breaking scale are assumed to be
the usual standard-model fields, together with the gauge-singlet fermion $s$
and two electroweak singlet scalar fields $\phi_j$ transforming under $G_\nu$
as
\eq
\phi_1 \sim ( -\sfrac 12, -\sfrac 12) \qquad\hbox{and}\qquad
\phi_2 \sim (0, -\sfrac 23).
\eeq
The lowest-dimension gauge- and $G_\nu$-invariant operators in the effective
lagrangian at this scale that contribute to the neutrino mass matrix are
\eq
\label\themodelII
\eqalign{
\hbox{dimension 5:} \quad & \frac {g_e}{M} \, (L_e H)(L_\mu H),
\quad \frac {g_\tau}{M} \, (L_\mu H)(L_\tau H); \cr
\hbox{dimension 6:} \quad & \frac {a_j}{M^2} \, (L_j H) \, s \phi_1^2,
\quad \frac {c}{M^3} \, ss \phi_2^3. \cr}
\eeq
Contributions to the remaining terms in the neutrino mass matrix are further
suppressed relative to \themodelII\ by additional powers of $M^{-1}$.

Replacing the scalars with their vacuum expectation values $\avg{H}=v$,
$\avg{\phi_j}=w_j$, and defining $g\equiv\sqrt{g_e^2+g_\tau^2}$, we find that
the heavy scale must be $M = (1 \times 10^7) gv$, and that the dimensionless
mass
parameters of eq.\ \massmatrix\ are
\eq
\alpha_j = \frac{a_j w_1^2}{g Mv}, \quad
\gamma = \frac{c w_2^3}{g M v^2}, \quad
\beta, \epsilon, \eta \ll \alpha, \gamma.
\eeq
In this case, the light neutrino states have masses
\eq
m_{\ell\pm} = \frac{m_{17}}2 \; \left[
\sqrt{\gamma^2 + 4{\alpha_1'}^2} \pm \gamma \right],
\eeq
with
\eqa
\Delta m_\ell^2 & = m_{17}^2 \gamma \sqrt{\gamma^2 + 4{\alpha_1'}^2}, \eol
\sin^2 2\theta_\ell & = \frac{4{\alpha_1'}^2}{\gamma^2 + 4{\alpha_1'}^2}. \eeol
\eeq
The splitting of the heavy neutrino states is negligible in this model.

MSW oscillations of the light states can be accomodated if we choose
\eg\ $g = 1$, $a_1 = 0.2$, $a_2 = 1$, and $c = 1$.
This gives $\alpha'_1 = 1 \times 10^{-8}$, $\alpha'_2 = 1 \times 10^{-7}$,
$\gamma = 1 \times 10^{-7}$, and
\eqa
\Delta m_\ell^2 & = 3 \times 10^{-6} \eV^2, \eol
\sin^2 2\theta_\ell & = 4 \times 10^{-2}. \eeol
\eeq

The two majorons of this model may be defined to couple to the charges
$Q_1 \equiv L - S$, $Q_2 \equiv L$, respectively, giving a lifetime for the
17 keV state of
\eq\label\lifetime
\tau = \frac{16\pi w_1^2}{{\alpha_2'}^2 m_h^3}.
\eeq
For the choice of parameters given above, the lifetime is $\sim 10^{10}$ sec.
This is in conflict with the cosmological structure formation bounds, but is
compatible with all other bounds.

The model considered above can be modified to accomodate the atmospheric
neutrino anomaly by adding a third electroweak singlet scalar transforming
under $G_\nu$ as
\eq
\phi_3 \sim (\sfrac 12, -\sfrac 12).
\eeq
Then there is an additional dimension 6 operator in the weak-scale effective
lagrangian:
\eq
\frac b{M^2} \; (L_\mu H) s \phi_3^2,
\eeq
which gives
\eq
\beta = \frac{b w_3^2}{gMv}.
\eeq

This model is nonminimal, in the sense that there are now more scalar fields
than order parameters.
However, it can easily accomodate the atmospheric neutrino anomaly for $w_3$
near the weak scale.
Because of the additional freedom in this model, it can also give rise to very
short heavy neutrino lifetimes.
For example, if we choose $w_1 = v / 2$, $w_2 = v$, $w_3 = 30v$,
$g = 0.1$, $a_1 = 0.1$, $a_2 = 0.01$, $b=1$, and $c = 0.01$,
we find that the model incorporates the MSW effect and atmospheric
neutrino oscillations, and the neutrino lifetime is
\eqa
\tau & = \frac{16\pi w_3^2}{m_h^3\beta^2}\eol
& \simeq 2 \times 10^{3} \sec, \eeol
\eeq
in the limit that $w_3 \gg w_1, w_2$ and $\beta \gg \alpha_i$.\footnote{
$^\dagger$}{\def\s{\scriptstyle} \eightrm It is also possible to choose
scalar quantum numbers so that $\beta$ is naturally much larger than the  other
elements of $\delta M$, and so have MSW and atmospheric oscillations,  short
lifetimes, and no VEVs below the weak scale. If, for example, $\s\phi_1 \sim
(1/3,-2/3)$, $\s\phi_2 \sim (-1/3, -1/3)$, and $\s\phi_3 \sim (2/3, -1/3)$ then
$\s\alpha_i \sim \gamma \sim O(v^2/M^2)$ while $\s\beta \sim O(v/M)$.}

\noindent$\bullet$ {\it Renormalizable Model}

As an example of a renormalizable model which can give rise to the two-scalar
effective theory discussed above, we add several gauge-singlet Dirac fermions
transforming under $G_\nu$ as
\eq
N_{1}^{\pm} \sim \pm(\sfrac 12,-\sfrac 12),
\qquad N_{2}^{\pm}  \sim \pm(0, \sfrac 13),
\qquad N_{3,4}^{\pm}  \sim \pm(1, 0),
\eeq
(Again, two copies of the last state are required to avoid accidental
symmetries of the neutrino mass matrix.)

The most general renormalizable interactions of this model are the usual
standard model interaction, with the addition of
\eq\eqalign{
\hbox{dimension 3:} \qquad & M_j N_j^+ N_j^-, \cr
\hbox{dimension 4:} \qquad & (L_j H) \, N_{3,4}^-,
\quad (L_\mu H) N_{3,4}^+, \quad
s N_1^+ \phi_1, \qquad s N_2^- \phi_2, \cr
& N_{3,4}^+ N_1^- \phi_1,
\quad N_{3,4}^+ N_1^+ \phi_1^*, \quad
 N_2^+ N_2^+ \phi_2, \quad N_2^- N_2^- \phi_2^*. \cr}
\eeq
If we assume that all dimensionless couplings are of order $0.1$--$1$, then we
obtain the effective theory presented above after integrating out the Dirac
fermions.

\section{Conclusion}

We have shown that several recently reported experimental anomalies in the
neutrino sector can be accounted for in a simple class of models with a single
light electroweak singlet fermion $s$, and an approximate
$G_\nu \equiv U(1)_{e-\mu+\tau} \times U(1)_s$ symmetry.
All neutrino mass hierarchies are understood in terms of the pattern in which
this symmetry is broken.
We have examined several models which can give rise to interesting symmetry
breaking patterns, and we always find that $G_\nu$ is broken near the weak
scale, a feature which we find very attractive. The models are compatible with
all astrophysical and cosmological bounds at present.

There are several ways in which the class of models we have discussed will be
experimentally probed in the forseeable future. The first and most obvious is
the ongoing effort to confirm or disprove the experimental anomalies which are
the motivation for these models. Second, solar neutrino oscillations are into a
sterile component, which should be detectable once neutral-current solar
neutrino events are observed, for example at SNO. Third, in models where the
majorons arise from scalar fields, they can contribute a large invisible width
to the Higgs via its decay to two majorons. This gives rise to observable
missing-energy events at LEP II and LHC or SSC for a large portion of
parameter space \JerryMarkus. Fourth, if we are fortunate
enough to observe another nearby supernova with detectors that count neutral
current events, and if the \simp\ lifetime is less than $10^4$ sec, then all
mu and tau neutrinos can have decayed before reaching the earth. Besides
depleting the $\nu_\mu$ and $\nu_\tau$ fluxes this could also prolong the
$\nu_e$ signal.
If, on the other
hand, the lifetime should be on the order of $10^9 - 10^{11}$ secs such a
neutrino may have interesting applications for galaxy formation.

\section{Acknowledgements}

\lblfoot\
This research was partially funded by funds from the N.S.E.R.C.\ of Canada and
les Fonds F.C.A.R.\ du Qu\'ebec.

\listrefs
\bye